\documentclass[aps,twocolumn,pra]{revtex4-1}

\usepackage{graphicx}
\usepackage{amsmath,mathtools}
\usepackage{color}
\usepackage{xcolor}
\usepackage{ulem}

\usepackage[hidelinks]{hyperref}

\begin{document}

\title{Noise induced coherent ergotropy of a quantum heat engine}

\author{Manash Jyoti Sarmah}

\author{Himangshu Prabal Goswami}
\email{hpg@gauhati.ac.in}
\affiliation{Department of Chemistry, Gauhati University, Jalukbari, Guwahati-781014, Assam, India}
\date{\today}

\begin{abstract}
We theoretically identify the noise-induced coherent contribution to the ergotropy of a four-level quantum heat engine coupled to a unimodal quantum cavity. We utilize a protocol where the passive state's quasiprobabilities can be analytically identified from the population-coherence coupled reduced density matrix. The reduced density matrix elements are evaluated using a microscopic quantum master equation formalism. Multiple ergotropies within the same coherence interval, each characterized by a positive and pronounced coherent contribution, are observed. These ergotropies are a result of population inversion as well as quasiprobability-population inversion, controllable through the coherence measure parameters. The optimal flux and power of the engine are found to be at moderate values of ergotropy with increasing values of noise-induced coherence. The optimal power at different coherences is found to possess a constant ergotropy. 

\end{abstract}

\maketitle

\section{Introduction}
The evolution of a quantum system can be influenced by the interaction of its discrete energy levels with external incoherent or stochastic fields, such as white-noise radiation, harmonic states, thermal radiation, or squeezed states of light\cite{Rotter_2015_review_qs_th_exp, PhysRevLett.55.2409_Squeezed_st_exp,ourjoumtsev2011observation_Sq_exp}. These external fields facilitate the coupling of populations and quantum coherences, allowing Fano interferences that contribute to the system dynamics\cite{doi:10.1073/pnas.1110234108_Qhe_power_increased_noise-induced_coherence, PhysRevResearch.3.013295Steady-state_Fano_coherences_in_a_V-type}. Such an interaction opens up non-conventional pathways in the form of specific superpositioned states, which can either enhance or suppress existing energy transfer processes in open and nonequilibrium quantum systems\cite{PhysRevA.88.013842_hpg01, PhysRevResearch.3.013295Steady-state_Fano_coherences_in_a_V-type, doi:10.1073/pnas.1110234108_Qhe_power_increased_noise-induced_coherence, PhysRevA.107.052217_sarmah_01, Chin_Huelga_Plenio_2012_orgin_of_long_induced_coherence,ps2_Wang:22}. Such dynamics are usually referred to as noise-induced coherence-assisted dynamics. 
Noise-induced coherence is instrumental in detecting disruptions in the optical path of probe photons by generating interference signals within a Mach–Zehnder interferometer integrated with a double-pass-pumped spontaneous parametric down-conversion process\cite{Kim_2024_interferometer} leading to significant enhancement in power output of lasers and solar cells. Noise-induced coherence has also been demonstrated to be responsible for highly efficient energy transfer in photosynthetic systems, a phenomenon that has been experimentally confirmed in polymer solar cells\cite{Daryani:23_noise-induced_coherencesolar_cells}. Significant enhancement in flux and power output in quantum heat engine prototypes and nitrogen vacancy-based quantum heat exchangers are also reported\cite{PhysRevE.97.042120_laser_power_coherence_in,doi:10.1073/pnas.1110234108_Qhe_power_increased_noise-induced_coherence, PhysRevResearch.4.L032034_Coherence-enhanced-q-dot, Liu2019_CoherentQuantumControl_NVC}.

The intriguing capability of noise-induced coherence to influence photophysical properties positions it as a compelling subject for investigating its impact on energy and particle exchange within nonequilibrium systems from a quantum thermodynamic perspective\cite{PhysRevA.107.052217_sarmah_01, PhysRevA.88.013842_hpg01, PhysRevA.86.043843_Rahav_Reducued_DM_1}.  In the realm of quantum thermodynamics, the quantum system's state plays a pivotal role in defining the relationships among thermodynamic variables at play. In the context of quantum energy exchange devices like engines or batteries, the presence of non-Gibbsian states enables the extraction of multiple work values during energy transfer processes\cite{A.E.Allahverdyan_2004_Maximalworkextractionfromfinitequantumsystems}. The ergotropy, a key metric quantifying the maximum work extractable from a quantum system, is determined by optimizing the full quantum space of cyclic unitary evolutions that allow transitions between active and passive states of the system\cite{Biswas2022extractionof, PhysRevE.102.042111_Ergotropy_from_coherences,e24060820_Quantum_Battery_Charged, PhysRevLett.122.210601_Dissipative_Charging_Quan_Battery}. This metric has been experimentally assessed in microscopic engines coupled to external loads\cite{Biswas2022extractionof, PRXQuantum.4.020309_zr_ergo, van2020single}.\par

Systems evolving through coupled populations and coherences offer a fertile ground for exploring the principles of quantum thermodynamics\cite{PhysRevE.102.042111_Ergotropy_from_coherences,doi:10.1126/science.1078955_Scully_Single_Heat_Bath, Latune2021quantum-coherences-in-thermal-machines, PhysRevA.104.062210-Suppressing-coherence, PhysRevLett.125.180603-Quantum-Coherence-and-Ergotropy}. A method has been proposed to evaluate the coherent contribution to ergotropy while considering the absence of coherence monotones, thus providing newer quantitative insights in the heat-to-work conversion in quantum systems\cite{PhysRevLett.125.180603-Quantum-Coherence-and-Ergotropy}. 
The primary objective of this study is to investigate how noise-induced coherence impacts the ergotropy of an extensively researched quantum heat engine prototype. We quantify the engine's dynamics through the microscopic equations of motion that couple the populations and coherences in the reduced density matrix\cite{Harbola_2012Reducued_DM_2, PhysRevA.86.043843_Rahav_Reducued_DM_1}.  The extent of influence of the noise-induced coherence is parameterized in the form of two system variables. Such parametrization of the noise-induced coherence is extremely common and widely used\cite{PhysRevA.86.043843_Rahav_Reducued_DM_1, doi:10.1073/pnas.1110234108_Qhe_power_increased_noise-induced_coherence, PhysRevA.88.013842_hpg01, PhysRevResearch.4.L032034_Coherence-enhanced-q-dot, PhysRevA.107.052217_sarmah_01, PhysRevE.97.042120_laser_power_coherence_in}. Optimization of the flux and power of many popular quantum engines have been performed using such parameters\cite{PhysRevA.88.013842_hpg01, PhysRevA.86.043843_Rahav_Reducued_DM_1, PhysRevA.107.052217_sarmah_01, doi:10.1073/pnas.1110234108_Qhe_power_increased_noise-induced_coherence}. These parameters allow us to monitor the role of noise-induced coherence in affecting the ergotropy. We show how the two parameters can be controlled to achieve population inversion, a necessary condition for observing finite ergotropy. \par
The paper is organized as follows. Firstly, in Sec.(\ref{model}) we introduce the popular engine and present the necessary equations of motion. In Sec.(\ref{sec-erg}), we discuss the protocol to evaluate the ergotropy by calculating quasiprobabilities using the equations of motion for the probabilities and coherences. In this section, we also analytically identify the conditions that lead to coherence-enabled population inversion resulting in multiple values of ergotropies after which we conclude.   


\section{The Engine}

\label{model}
\begin{figure}[t!]
\centering
\includegraphics[width = 9cm]{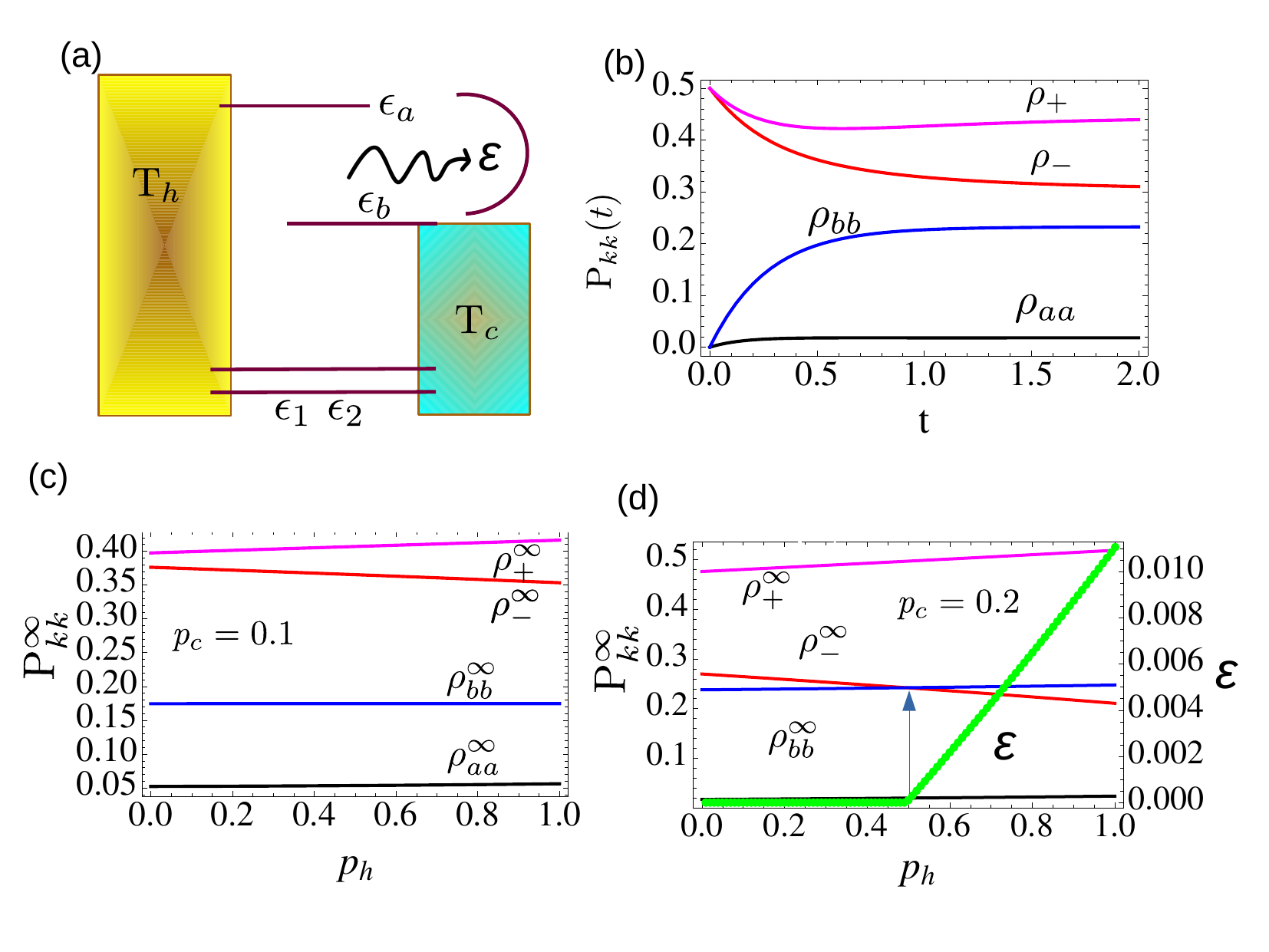}
\caption{\label{fig1}
(a) Schematic representation of the QHE where the four levels are coupled to two thermal reservoirs
and a unimodal cavity in an asymmetric arrangement. The ergotropy ${\cal E}$ is evaluated for the work mode at $\epsilon_a-\epsilon_b$. (b) Time evolution  quasiprobabilities, $P_{\pm}(t)$ and the populations $\rho_{a,b}$ obtained from a quantum master equation. At each point of time, the passive state is $\hat P_\rho=diag\{\rho_+,\rho_-,\rho_{bb},\rho_{aa}\}$. No ergotropy is obtainable from this state even though the active and passive states are different. (c), (d) Steadystate value of $P_{kk}^\infty$ as a function of the hot coherence parameter $p_h$ at cold coherence parameter $p_c =0.1$ and $0.2$ respectively. In (d), the ergotropy scale is shown in the RHS axis-pane. It becomes finite after $p_h=0.49$ (shown as an upward arrow). Engine parameters are fixed at $\epsilon_1=\epsilon_2=0.1,\epsilon_b=0.4,\epsilon_a=1.5, g =r=1,T_c=0.5$ and $T_h =2$. Ergotropy has the same units if energy and $p_h,p_c$ are dimensionless. Natural units are used ($\hbar =k_B=1$) throughout. }
\end{figure}

The quantum heat engine (QHE) model comprises four quantum levels that are asymmetrically connected to two harmonic reservoirs, with the upper two levels linked to a unimodal cavity, as depicted schematically in Fig. \ref{fig1}(a). Experimentally, QHE with similar configurations have been successfully implemented using cold Rb and Cs atoms in magneto-optical traps \cite{zou2017quantum,bouton2021quantum}. The total Hamiltonian of the four-level QHE is $\hat{H}_{T}\,=\,\sum_{m\,=\,1,2,a,b}
        \epsilon_{m}|m \rangle\langle m |+\hat H_\ell+\hat H_\nu+\hat{V}_{sb}+\hat V_{s\ell}$, with the system-reservoir and system-cavity coupling Hamiltonians given by,
$
\hat V_{sb}=\sum_{k\,\in\nu=h,c}\sum_{i\,=\,1,2}\sum_{\alpha=\,a,b}r_{ik}\hat{a}_{\nu k}|\alpha\rangle\langle i|+h.c$ and $ 
\hat V_{s\ell}=g\hat{a}_{\ell}^\dag|b\rangle\langle a|+h.c$. The system-reservoir coupling of the $i$th state with the $k$th mode of the reservoirs is denoted by $r_{ik}$ while $g$ is the strength of coupling between the cavity and the upper two states. $\hat H_\ell= \hbar\omega_\ell\hat{a}_{\ell}^{\dag}\hat{a}_{\ell}$ is the Hamiltonian for the cavity mode and $\hat H_\nu=\sum_{k}E_{\nu k}\hat{a}_{\nu k}^{\dag}\hat{a} 
        _{\nu k}$ is the Hamiltonian for the $\nu$-th reservoir. $E_{\nu k}$, $\hbar\omega_{\ell}$ and $\epsilon_{m}$ denote the bath's, the unimodal cavity, and the system's $m$th level energy respectively.   
The radiative decay originating from the transition $|a\rangle\to|b\rangle$ is the work done by the engine. The maximum work extractable in this mode is the ergotropy, ${\cal E}$, of the engine. This QHE has been thoroughly studied using a Markovian quantum master equation and we refer to the following works for the derivation of the master equation   \citep{PhysRevA.86.043843_Rahav_Reducued_DM_1, PhysRevA.88.013842_hpg01, Harbola_2012Reducued_DM_2,PhysRevE.99.022104_giri_ml, PhysRevA.107.052217_sarmah_01, sarmah2023learning}. The matrix elements of the reduced density matrix $\rho$ result in four populations, $\rho_{kk}, k = 1, 2, a,b$ coupled to the real part of a noise-induced coherence term, $\rho_{12}$, which arise due 
to interactions with the hot and the cold reservoirs so that $|\rho\rangle =\{\rho_{11},\rho_{22},\rho_{aa},\rho_{bb},\rho_{12}\}$.
The other quantum coherences between states $|a\rangle$ and $|b\rangle$ are decoupled from the rest of the elements and do not contribute to the populations, $\rho_{kk}$, or the noise-induced coherence, $\rho_{12}$. 
Under the symmetric coupling regime, the dynamics are given by a master equation of the type, $\dot{|\rho\rangle} = \mathcal{L}|\rho\rangle$ with ${\cal L}=$
\begin{small}
\begin{equation}
\label{Louv-eq}
\begin{pmatrix}
-\displaystyle rn&0&r\tilde n_h&r\tilde n_c&-ry\\
0&-\displaystyle r n&r\tilde n_h&r\tilde n_c&-ry\\
rn_h&rn_h& 
-(g^2\tilde n_\ell+2r\tilde n_h)&g^2 n_\ell &2rp_hn_h\\
r\tilde n_c&r\tilde n_c&g^2_{}\tilde n_\ell&-(g^2n_\ell+2r\tilde n_c)&2rp_cn_c\\
-ry/2&-ry/2&rp_h\tilde n_h&rp_c\tilde n_c&-rn
\end{pmatrix}\\[2mm]
\end{equation}
\end{small}
The Bose-Einstein distributions are denoted by $n_\nu$  and $n=n_c+n_h$, $y=n_cp_c+n_hp_h$  and $\tilde n_\nu=1+n_\nu$. The extent of coupling between populations and coherences is quantified by the presence of the terms $rp_\nu$, where $\nu = c, h$, in the coherence, $\rho_{12}$. The two dimensionless parameters $p_h$ and $p_c$  measure the strength of noise-induced coherence from the hot and the cold bath respectively and have been extensively explored \cite{doi:10.1073/pnas.1212666110Photosynthetic_reaction_center_as_a_qhe, doi:10.1073/pnas.1110234108_Qhe_power_increased_noise-induced_coherence, PhysRevA.88.013842_hpg01, Harbola_2012Reducued_DM_2, sarmah2023learning}. It is obtained by parametrizing terms of the type $|r_{1\nu}r_{2\nu}|^2$ in the perturbative equations of motion. Under symmetric coupling, we have $|r_{1\nu}|=|r_{2\nu}| = r$, and
$
 |r_{1\nu}r_{2\nu}|^2=r\sqrt{|\cos\phi_\nu|}$ with $\sqrt{|\cos\phi_{\nu}|}=p_\nu$, and $0 \le p_h, p_c \le 1$. We refer to the following works for a detailed discussion of these terms\cite{PhysRevA.86.043843_Rahav_Reducued_DM_1, doi:10.1073/pnas.1110234108_Qhe_power_increased_noise-induced_coherence, PhysRevA.88.013842_hpg01, PhysRevResearch.4.L032034_Coherence-enhanced-q-dot, PhysRevA.107.052217_sarmah_01, PhysRevE.97.042120_laser_power_coherence_in, sarmah2023learning}. A perpendicular orientation of
the individual dipole vectors kills the coherence, i.e., $p_c = p_h = 0$. A parallel orientation generates the maximum value
of the coupled term, $r$, and decouples the population from the coherences. The evolution of the reduced density matrix elements in the absence and presence of coherences are shown in Figs. (\ref{fig1}c and d) respectively.

\section{Ergotropy from quasiprobability}
\label{sec-erg}
Ergotropy, ${\cal E}(t)$, is the difference between two expectation values of the energy, one with respect to the active system density matrix $\hat \rho(t)$ and the other with respect to the passive system density matrix, $\hat P_\rho(t)$ \cite{PhysRevLett.125.180603-Quantum-Coherence-and-Ergotropy, A.E.Allahverdyan_2004_Maximalworkextractionfromfinitequantumsystems}. Mathematically,
\begin{align}
\label{eq-ergotropy}
 {\cal E}(t) &=\langle \hat H_o(t)\rangle-\langle \hat H_o(t)\rangle_P
\end{align}
where $\hat H_o$ is the Hilbert space system Hamiltonian arranged in the ascending eigenbasis, i.e $\hat H_o=diag\{\epsilon_1,\epsilon_2,\epsilon_b,\epsilon_a\}$. The active state, $\hat\rho(t)$ is the system density matrix written in the basis of $\hat H_o$. $\langle \hat H_o(t)\rangle_P$ is an expectation value obtained from a different density matrix, $\hat P_\rho (t)$, a particular passive state (in a rearranged or a renormalized basis) that generates the minimum expectation value of the operator $\hat H_o$ so that the difference in Eq. (\ref{eq-ergotropy}) is maximal. For diagonal density matrices, the passive state is readily identifiable by rearranging the elements in descending order.  In the considered QHE, the real part of the coherence couples to the populations of the states $|1\rangle$ and $|2\rangle$, and hence a direct identification of the passive state is not straightforward. For the current engine, we construct the nondiagonal Hilbert space density matrix by considering the population and coherences obtained from the reduced density matrix. Since, $\rho_{11}=\rho_{22}$ (degenerate) we have, 
\begin{equation}
\label{eq-pass}
\hat\rho(t)=
\begin{pmatrix}
\rho_{11}(t)&\rho_{12}(t)&0&0\\
\rho_{12}(t)&\rho_{11}(t)&0&0\\
0&0&\rho_{bb}(t)&0\\
0&0&0&\rho_{aa}(t)
\end{pmatrix}.
\end{equation}
The passive state, $\hat P_\rho(t)$ can constructed by
the  transformation $\hat P_\rho(t) =\hat U\hat\rho(t)\hat U^\dag$, where $\hat U$ is a cyclic unitary operator that first diagonalizes $\hat\rho(t)$ and rearranges the eigenbasis in descending order. The elements of $\hat P_\rho(t)$ are hence quasiprobabilities\cite{PhysRevLett.116.013603_quasi}. The top left block of Eq.(\ref{eq-pass}) is a Toeplitz type of symmetric matrix block whose eigenvalues are well known. For the current scenario, several passive states can be created by considering the set of eigenvalues of $\hat\rho (t)$, i.e. $\{\rho_-(t),\rho_+(t),\rho_{bb}(t),\rho_{aa}(t)\}$ with the quasiprobabilities $\rho_{\pm}(t)=\rho_{11}(t)\pm\rho_{12}(t)$. The evolution of the time-dependent eigenvalues, $P_{kk}(t)$, of $\hat\rho(t)$ are shown in Fig. (\ref{fig1}b) for a fixed set of engine parameters. The matrix elements $P_{kk}(t)$ correspond to populations when $k=a,b$ and quasiprobabilities when $kk=\pm$. Varying the Hamiltonian parameters changes the quasiprobabilities $\rho_{\pm}(t)$ as well as the populations. $\hat P_\rho(t)$ can be uniquely identified by rearranging the eigenvalues of $\rho(t)$ in the following fashion  $P_{kk}(t)\ge P_{k+1/k-1}(t)$ each time the Hamiltonian parameters are varied\cite{PhysRevE.102.042111_Ergotropy_from_coherences, PhysRevLett.125.180603-Quantum-Coherence-and-Ergotropy}. From  Fig. (\ref{fig1}b), the passive state for the chosen parameters can be readily identified to be  $\hat P_\rho(t)=diag\{\rho_+(t),\rho_{-}(t),\rho_{bb}(t),\rho_{aa}(t)\}$ by looking at the curves from top to bottom. Using this definition of the passive state in Eq. (\ref{eq-ergotropy}), the time-dependent ergotropy can be obtained analytically, which is zero even though $\hat \rho(t)\neq \hat P_\rho(t)$. This is because the quasiprobabilities in the passive state are symmetrically displaced eigenvalues of the active state.   Passive states that do not yield a finite ergotropy are Gibbs or thermal states. Hence the passive state $diag\{\rho_+(t),\rho_{-}(t),\rho_{bb}(t),\rho_{aa}(t)\}$,  though composed of quasiprobabilities can still be regarded as Gibbsian since no work can be extracted from that state irrespective of any engine parameter.

We next focus on the steadystate ergotropy, ${\cal E}$ which is obtained by taking the steadystate values of the populations and coherence in Eq. (\ref{eq-ergotropy}). The steadystate values, $\rho_{kk}^\infty$ and $\rho_{12}^\infty$ are obtained by setting $|\dot\rho\rangle=0$. We denote the steadystate values with the superscript $^\infty$, omit the $t$ dependence, and keep the rest of the notation of the density matrix elements as before. We begin by evaluating $P_{kk}^\infty$ as a function of the hot bath coherence parameter $p_h$ for a set of engine parameters as shown in Figs. (\ref{fig1}b and c). For the entire range of $p_h$, the passive state remains constant, $diag\{\rho_+^\infty,\rho_{-}^\infty,\rho_{bb}^\infty,\rho_{aa}^\infty\}$. Therefore the ergotropy is zero for the entire range of the noise-induced-coherence parameter $p_h$. For the same engine parameters, changing the cold noise-induced coherence parameter from $p_c=0.1$ to $0.2$, finite ergotropy is observed from $p_h=0.5$ onwards which increases linearly as seen in Fig. (\ref{fig1} d). The finite steadystate ergotropy is because the passive state changes at the point $p_h =0.5$ from being $diag\{\rho_+^\infty,\rho_{-}^\infty,\rho_{bb}^\infty,\rho_{aa}^\infty\}$ to $diag\{\rho_+^\infty,\rho_{bb}^\infty,\rho_{-}^\infty,\rho_{aa}^\infty\}$ as a result of inversion between the quasiprobability ($\rho_{-}^\infty$) and the population $\rho_{bb}^\infty$. Substituting the latter definition of the passive state in Eq. (\ref{eq-ergotropy}), we obtain the steadystate ergotropy, in this case, to be ${\cal E}=\epsilon_a (\rho_{bb} ^\infty- \rho_{aa}^\infty) + \epsilon_1 (\rho_{11}^\infty - \rho_{bb}^\infty - \rho_{12}^\infty) + \epsilon_b(\rho_{aa}^\infty + \rho_{12}^\infty-\rho_{11}^\infty)$.

\begin{figure}[b!]
\centering
\includegraphics[width = 9cm]{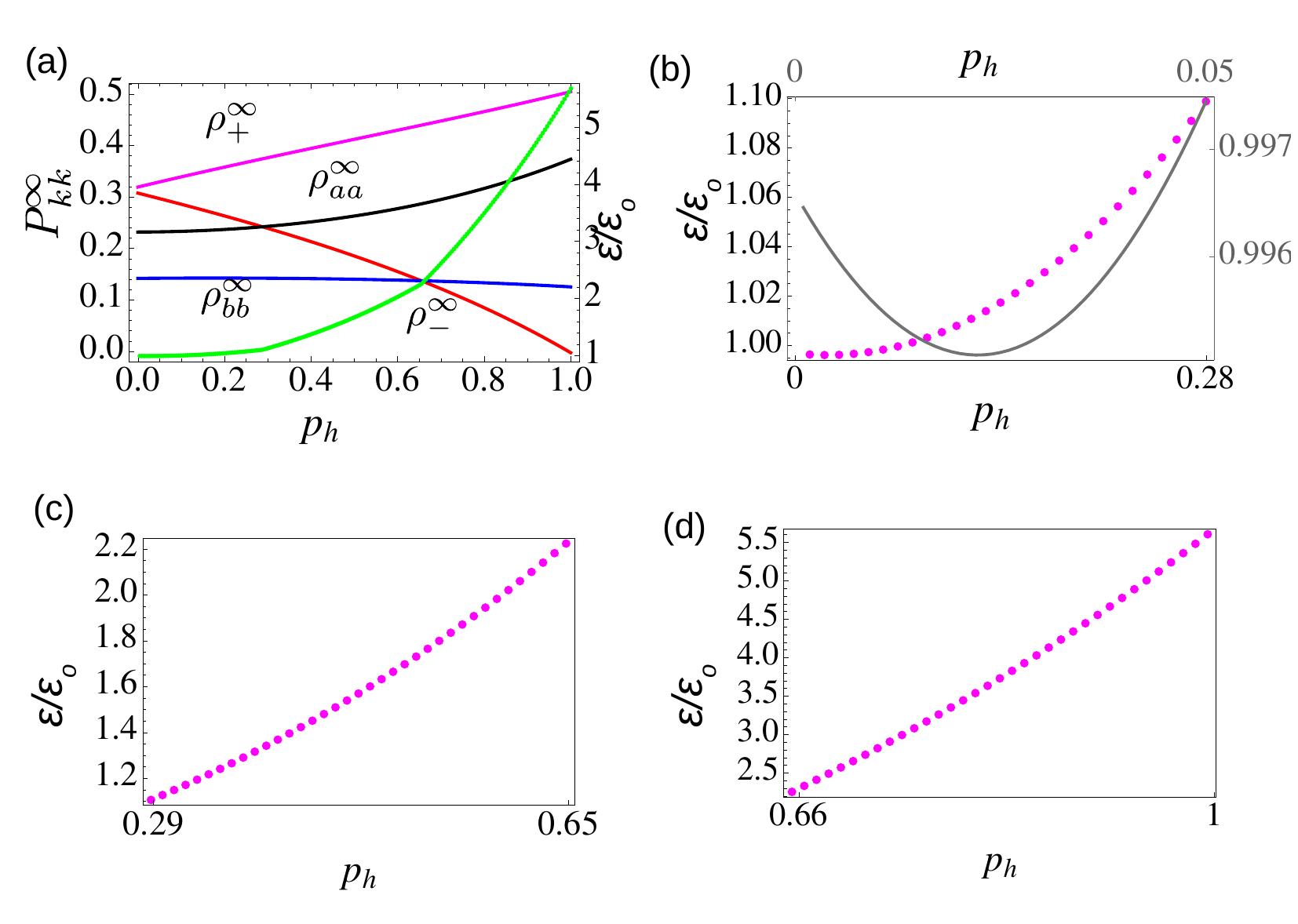}
\caption{\label{fig2}
(a) Crossover between three different passive states as  $p_h$ increases leading to multiple ergotropies, each shown separately for different intervals of $p_h$ as dotted lines in (b), (c) and (d) for $T_h=5$ and $p_c=0.1$ (scaled by incoherent ergotropy ${\cal E}_o$). In (b) ${\cal E}/{\cal E}_o<0$ is represented by the solid line ($0\le p_h<0.05$). Other parameters are the same as Fig. (1). Ergotropy has the same units if energy. Natural units ($\hbar=k_b\to 1$) are employed. }
\end{figure}
Within the same hot coherence interval, there can be multiple ergotropies present as the engine parameters are changed. This is shown in Fig. (\ref{fig2}). In Fig. (\ref{fig2}a), we plot the density matrix populations, $P_{kk}^\infty$ for a different set of engine parameters as a function of $p_h$. We see that there are three different passive states. Each passive state is characterized by a crossover between the matrix elements $P_{kk}$. From $p_h=0$ to $p_h=0.28$, the passive state in this coherence interval  is $diag\{\rho_+^\infty,\rho_{-}^\infty,\rho_{aa}^\infty,\rho_{bb}^\infty\}$.
and the ergotropy is given by ${\cal E}=\epsilon_1(\rho_{aa}^\infty-\rho_{bb}^\infty)$. ${\cal E}_o$ is the ergotropy in the absence of coherences (i.e. when $p_c=p_h=0$, a classical ergotropy analog or an incoherent ergotropy). The ratio ${\cal E}/{\cal E}_o$ is overlayed in the same graph with rescaled axes (shown in r.h.s of the frame). The quantity is a measure of the influence of the coherent contribution to the total ergotropy. ${\cal E}/{\cal E}_o>(< 1)$ corresponds to the total ergotropy being greater (less) than the ergotropy in the absence of the noise-induced coherences. Note that coherent contribution to ergotropy has been proposed where it has been defined as the difference between the ergotropies in the presence and absence of coherences. We take it as a ratio since it allows us to estimate the manifold increase in comparison to incoherent ergotropy. This ratio within the region $0< p_h\le 0.28$ is also shown in Fig. (\ref{fig2}b). It is seen to be nonlinear (dotted line). Near $p_h=0$, the total ergotropy is less than the incoherent ergotropy (${\cal E}/{\cal E}_o<1$, solid line). It then increases to go beyond the incoherent ergotropy by $10\%$. 

In the next coherence interval, $0.28<p_h\le 0.65$, the passive state changes to $\hat P_\rho =diag\{\rho_+^\infty,\rho_{aa}^\infty,\rho_{-}^\infty,\rho_{bb}^\infty\}$ from $diag\{\rho_+^\infty,\rho_{-}^\infty,\rho_{aa}^\infty,\rho_{bb}^\infty\}$. The ergotropy is ${\cal E}=(\epsilon_1 - \epsilon_b) (\rho_{11}^\infty - \rho_{aa}^\infty - \rho_{12}^\infty)$. The ergotropy in this region is higher than in the previous region with a nonlinear increase that is more than twofold (Fig.\ref{fig2}c). In the coherence interval  $0.65\le p_h\le 1$, the passive state again changes to  $\hat P_\rho =diag\{\rho_+^\infty,\rho_{aa}^\infty,\rho_{bb}^\infty,\rho_{-}^\infty\}$ such that the ergotropy is ${\cal E}=\epsilon_b(\rho_{aa}^\infty - \rho_{bb}^\infty ) + \epsilon_{1} (\rho_{11}^\infty - \rho_{aa}^\infty - \rho_{12}^\infty) + \epsilon_a(\rho_{aa}^\infty + \rho_{12}^\infty-\rho_{11}^\infty)$.
The ergotropy ratio here rises much more sharply than in the previous two cases increasing nonlinearly to beyond a fivefold value (Fig. (\ref{fig2}d)).

\begin{figure}
\centering
\includegraphics[width = 9cm]{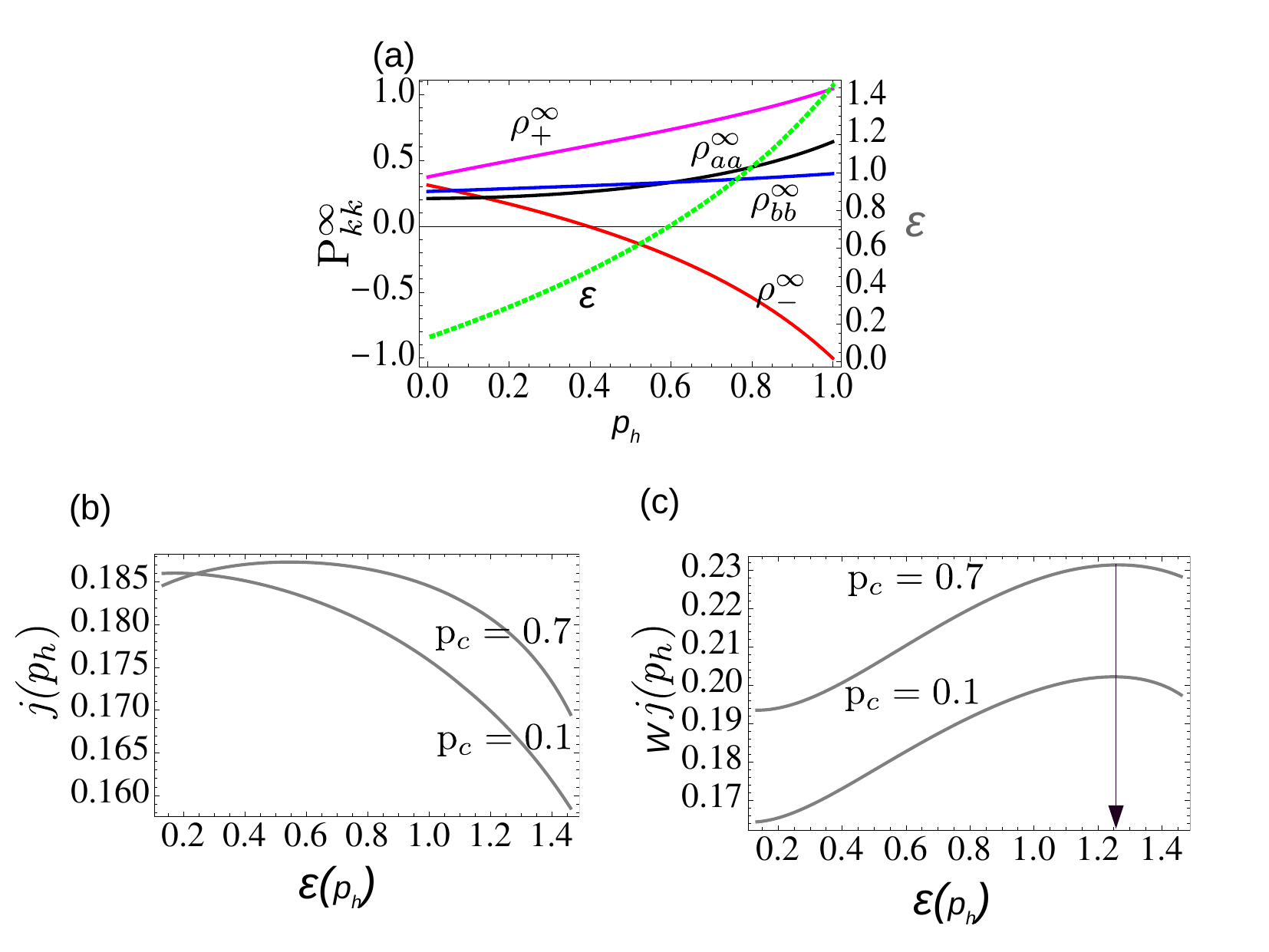}
\caption{\label{fig3} (Color online)
(a) Positive ergotropy for negative quasiprobability, $\rho_-$ starting at $p_h=0.38$ for $p_c=0.6$. The ergotropies for each passive state existing in a coherence interval ($p_h\in[0,0.05),[0.05,0.14), (0.14,0.62]$ and $(0.62,1]$) are positive as seen from  the  monotonously increasing (green) curve (scale on the right pane). (b) Variation of the coherent flux $j(p_h)$ into the work mode as a function of the ergotropy evaluated in the interval $0\le p_h\le 1$. Note the trade-off between the maximal values of the flux and ergotropy. (c) Variation of the engine's output power $Wj(p_h)$ into the work mode as a function of the ergotropy evaluated in the interval $0\le p_h\le 1$. The arrow points to the same value of ergotropy at maximum power. Parameters are the same as in Fig. (1). Ergotropy has the same units if energy and natural units ($\hbar=k_b =  1$) are employed.}
\end{figure}

The shuffling of the engine's ergotropies as any Hamiltonian parameter is varied is related to the crossover between the matrix elements $P_{kk}^\infty$. Such crossovers are a direct consequence of population inversion or an inversion between quasiprobability and populations. Any parameter that can be tuned to achieve such inversion will alter the system ergotropy. When it comes to the coherence parameters, $p_c$ and $p_h$, not all inversions are possible.
 In the passive state's precursor density matrix, $diag\{\rho_+^\infty,\rho_{aa}^\infty,\rho_{bb}^\infty,\rho_{-}^\infty\}$, it is not possible to make an inversion of the type $\rho_+<\rho_k, k =-, aa,bb$ by tuning the coherences alone. $\rho_{+}>\rho_{-}$, since both its constituent terms, $\rho_{11}$ and $\rho_{12}$, are positive. Hence $\rho_{+}$ shall always be the first element of the passive density matrix. It is possible that $\rho_-<0$. Observation of such negative quasiprobability is not new. Diagonalization of the population-coherence coupled density matrices is known to allow such observations, e.g. in counting statistics when interference terms get separated from their classical counterparts in the Hilbert space, one observes negative quasiprobabilities \cite{PhysRevLett.116.013603_quasi}. However, such negative quasiprobabilities do not result in the ergotropy being negative. When, $\rho_-<0$, $\rho_-$ is, by default, the last element of the passive state's density matrix. Such a condition can be triggered by tuning the noise-induced coherences alone. For example, in the limit of $n_c\ll n_h $ and $g=r = 1$, $\rho_-$ is negative when 
$p_h> p_h^o$ with
\begin{align}
 p_h^o&=\frac{\sqrt{n_1}-(2 n_h (n_\ell+2)+n_\ell (p_c+3)+p_c+5)}{2 \tilde n_h
   (n_\ell+2)}.
\end{align}
Here, $n_1=8 n_h^2 (n_\ell+2)^2+8 n_h (n_\ell+2) (3 n_\ell+5)+17 n_\ell^2+(n_\ell+1)^2 p_c^2+2
   \tilde n_\ell^2 p_c+58 n_\ell+49$. 
The passive state's density matrix, which has a form, $\hat P_{\rho}=diag\{\rho_+^\infty,\rho_{bb}^\infty,\rho_{aa}^\infty,\rho_{-}^\infty\}$ yields an ergotropy, ${\cal E}=(\epsilon_a-\epsilon_1)(\rho_{bb}^\infty-\rho_-^\infty)$, a positive quantity. This is highlighted in Fig. (\ref{fig3}a) in the coherence interval $0.38 \le p_h \le 1$ where $\rho_{-}^\infty<0$ and ${\cal E}>0$. The ergotropy scale is shown in the r.h.s pane of the figure axis.

Under the same conditions, a passive state density matrix of the type, $\hat P_{\rho}=diag\{\rho_+^\infty,\rho_{aa}^\infty,\rho_{bb}^\infty,\rho_-^\infty\}$ can also exist and yields an ergotropy, ${\cal E}=\epsilon_b (\rho_{aa}^\infty - \rho_{bb}^\infty) - \epsilon_1 (\rho_{aa}^\infty - \rho_{-}^\infty ) + \epsilon_a (\rho_{bb}^\infty -\rho_{-}^\infty)$.
The first and last terms in the expression are always positive. The second term is always negative but its magnitude is less than the first and third terms, resulting in an overall positive ergotropy. Although not simple to show analytically, the conditions under which the mathematical expression for the engine's ergotropy becomes negative do not belong to the engine's coherence interval of $0\le p_c,p_h\le 1$.

Interestingly, the switch between the two ergotropies when $\rho_-^\infty< 0$ is solely due to the population inversion between the states $|b\rangle$ and $|a\rangle$. This is an ideal situation to study since population inversion between the upper two states makes the mode behave like a laser. This laser photon in that mode is equivalent to the work done by the engine\cite{PhysRevA.94.053859_harris_eit, PhysRevA.88.013842_hpg01}. Such work done via lasing has been experimentally demonstrated too\cite{PhysRevLett.119.050602_zou1, PhysRevA.103.062205_zou_dop}.
To create a population
inversion ($\rho_{aa}>\rho_{bb}$), the ratio between the Bose-Einstein factors of the two thermal baths must be bound in the following fashion,
%
$x_-(p_c,p_h)\le\frac{n_c}{n_h} \le x_+(p_c,p_h)$
where the upper ($x_-(p_c,p_h)$) and lower bound ($x_+(p_c,p_h)$) is given by,
\begin{eqnarray}
\label{popinveqn}
x_\pm(p_c,p_h) &\!=\!&\mathcal{B}(p_c,p_h)\pm
\sqrt{\mathcal{B}(p_c,p_h)^2-\frac{(g^2-2r)(1+p_h^2)}{(g^2+2r)(1+p_c^2)}}\nonumber\\
\end{eqnarray}
with the term $
\mathcal{B}(p_c,p_h)=\{g^2(1-p_c p_h)-r(p_h^2-p_c^2)\}/\{(g^2+2r)(1-p_c^2)\}.$
Since $n_c/n_h$ is real and positive, we construct the following two scenarios. Firstly, when,  $g^2 <  \{r(p_h^2-p_c^2)\}\{1-p_cp_h\}$, $\mathcal{B}(p_c,p_h)>0$ and $g^2<2r$. This leads to  $x_-(p_c,p_h)$ being less than zero and $x_+(p_c,p_h)$ being greater than zero. Here population inversion, i.e., $\rho_{aa}^\infty>\rho_{bb}^\infty$, takes place when $ n_c/n_h\le x_+(p_c,p_h)$. Secondly,  ${\cal B}<0$, i.e., $g^2 >  \{r(p_h^2-p_c^2)\}\{1-p_cp_h\}$. Here there are two more conditions under which population inversion can happen. If $g^2<2r$, then 
$n_c/n_h\le x_+(p_c,p_h)$ and for $g^2>2r$, we must maintain
$ x_-(p_c,p_h)\le n_c/n_h\le x_+(p_c,p_h)$.
Combining the two cases, we conclude that
if $g^2<2r$, population inversion between $|a\rangle$ and $|b\rangle$ is possible only 
if $n_c/n_h\le x_+(p_c,p_h)$, otherwise the necessary condition is 
$x_-(p_c,p_h)\le n_c/n_h\le x_+(p_c,p_h)$.
For the incoherent situation (no coherence effects), $x_-(p_c,p_h)<0$ so that the lasing
condition for inversion reduces to, 
$n_c/n_h\le (2r-g^2)/(2r+g^2)$ with  no possible inversion if $g^2>2r$.

It is an established and well-understood fact that the flux and power of coherent engines can be optimized beyond the classical or incoherent counterparts by tuning the noise-induced coherences\cite{PhysRevA.86.043843_Rahav_Reducued_DM_1, PhysRevA.88.013842_hpg01, Harbola_2012Reducued_DM_2,doi:10.1073/pnas.1212666110Photosynthetic_reaction_center_as_a_qhe, doi:10.1073/pnas.1110234108_Qhe_power_increased_noise-induced_coherence}. Both the flux, $j$, and the power, ${\cal P}$ of this particular QHE is known analytically. The flux is given by, $j=g^2(\tilde n_\ell\rho_{aa}^\infty-n_\ell\rho_{bb}^\infty)$\cite{PhysRevA.88.013842_hpg01}. And the power is, ${\cal P}=jW$. We take the standard definition of work $W=(\epsilon_a-\epsilon_b)-k_BT_c\log(\rho_{aa}^{\infty}/\rho_{bb}^{\infty})$\cite{doi:10.1073/pnas.1212666110Photosynthetic_reaction_center_as_a_qhe, doi:10.1073/pnas.1110234108_Qhe_power_increased_noise-induced_coherence}. We evaluate the flux, power and the ergotropy as a function of $p_h$. We plot $j(p_h)$ as a function of ${\cal E}(p_h)$ in Fig. (\ref{fig3}b) for the coherence interval $0 \le p_h \le 1$. $j(p_h)$ is not proportional to ${\cal E}(p_h)$. For different $p_c$ values, it increases nonlinearly to an optimal value and then decreases. However, in contrast  to the behavior of the flux as a function of $p_h$ \cite{PhysRevA.88.013842_hpg01}, the optimum value of the flux, $j(p_h)$  doesn't correspond to a maximum value of the ergotropy. Maximum ergotropy,${\cal E}(p_h)$ is when the hot coherence parameter is at its highest value of unity. At that value, the flux is found to be the lowest. This is a trade-off between the ergotropy and the flux, where flux increases up to a certain extent as the ergotropy increases but reduces as the ergotropy keeps increasing. In Fig. (\ref{fig3}c), we plot the power as a function of the ergotropy evaluated in the interval $0\le p_h\le 1$. Similar to the flux behavior, the highest value of power is at some moderate value of the ergotropy and not at its maximal value for different $p_c$ values. Interestingly, the ergotropy value for which the power is maximum remains constant for different values of the cold coherence parameter $p_c$ as shown by the downward-pointing arrow in Fig. (\ref{fig3}c) although the absolute value of power is larger for larger $p_c$.

\section{Conclusion}
We performed a thorough investigation of the role of noise-induced coherences on the ergotropy (${\cal E}$ ) of a four-level quantum heat engine coupled to a cavity. The identification protocol of the passive state required to evaluate the ergotropy was based on diagonalizing the Hilbert space density matrix in the presence of population-coherence coupling. The Hilbert space matrix elements were taken to be the reduced density matrix elements obtained from a quantum master equation. The influence of coherence (which is a result of the engine's states with the thermal baths) on the total ergotropy was evaluated by defining a ratio ${\cal E}/{\cal E}_o$ (${\cal E}_o$ a measure of the influence of the incoherent contribution to the total ergotropy) by varying the noise-induced coherence measure parameters. Multiple ergotropies were found to exist within the same coherence interval due to the cross-over between several passive states. For each passive state, we observed a positive ergotropy with ${\cal E}/{\cal E}_o$ being $>1$ for almost the entire range of hot coherence parameter. However, ${\cal E}/{\cal E}_o$ is contained below unity for very low values of the hot coherence parameter. We also show that ergotropy remains positive even in the presence of coherence-enabled negative quasiprobabilities in the passive state. Bounds on the ratio of the Bose-Einstein distribution factors of the two thermal baths were obtained for a coherence-enabled inversion to be feasible between the upper two states of the engine, $|a\rangle$ and $|b\rangle$, which results in positive ergotropy in the presence of coherence-enabled negative quasiprobabilities. Within the same coherence interval, we observed that the flux and power are not proportional to the ergotropy. The optimized value of the flux at the optimal value of coherence does not align with maximal ergotropy. Rather, the flux peaks at moderate values of the ergotropy. High (low) flux values were observed at low (high) ergotropies. Likewise, the highest power was found to occur at intermediate ergotropy values. However, high (low) power values were observed at high (low) ergotropies. Irrespective of the value of the cold coherence parameter the ergotropy value for which the power is maximum remains constant in a hot coherence interval. 

\bibliography{references}

\end{document}